\documentclass{bms}
\usepackage{lipsum}

\usepackage{graphicx,amsmath,bm}

\begin{document}

\title{Effect of short-ranged spatial correlations on the Anderson localization of phonons in mass-disordered systems}


\author{Wasim Raja Mondal\textsuperscript{1}, \and N.\ S.\ Vidhyadhiraja\textsuperscript{1, *}}

\affilOne{\textsuperscript{1}Theoretical Sciences Unit, Jawaharlal Nehru Centre for Advanced Scientific Research, Bangalore 560 064, India.}


\twocolumn[{
\maketitle
\begin{abstract}
We investigate the effect of spatially correlated disorder on the Anderson transition of phonons in three dimensions using a Green’s function based approach, namely, the typical medium dynamical cluster approximation (TMDCA), in mass-disordered systems. We numerically demonstrate that correlated disorder with pairwise correlations mitigates the localization of the vibrational modes. A correlation driven localization-delocalization transition can emerge in a three-dimensional disordered system with an increase in the strength of correlations. 
\end{abstract}
\keywords{Anderson localization; Phonon localization.}
}]
\corres{raja@jncasr.ac.in}
\vspace{-36pt}
\section{Introduction}
Anderson introduced an ideal theoretical model containing the essential ingredients for studying the nature of one-electron states in disordered systems [1]. The model assumed non-interacting electrons moving through a lattice and allowed to hop only to nearest-neighbor sites. Disorder was introduced in the local orbital energies, which were independent quenched random variables distributed according to some specified probability distribution. Anderson predicted that the wave function may become exponentially localized with a characteristic localization length depending on the strength of disorder. Scaling theory[2] bolstered Anderson's idea of localization[1] by considering non-interacting electron systems with uncorrelated disorder. It found that all one-electron states are exponentially localized in one and two dimensions even for infinitesimal amount of disorder, with a true metal-insulator transition occurring
only in three dimensions (3D) whence the single-particle states may survive as extended states for weak disorder. A series of analytical, numerical and experimental results find strong agreement with one parameter
scaling theory of localization. However, the characteristics of the disorder potential can have a strong impact on Anderson localization. In particular, spatial correlations in the disorder can markedly change the conventional physics of Anderson localization. 

Such correlated disorder is relevant to transport properties of binary solids, DNA[3,4], graphene [5], quantum Hall wires[6], topological insulators[7] and so on. Recently, there has been a growing interest in understanding the effect of spatial correlations on Anderson localization due to tremendous experimental progress. Clement et al.[8] developed a experimental technique for creating correlated disorder through the laser speckle method. In this method, one can accurately control the spatial correlation length. A spatial correlation induced localization-delocalization transition has been experimentally observed in GaAs-AlGaAs superlattices[9]. Very recently, a transition between algebraic localization and delocalization in a 1D disordered potential with a bias has been reported[10]. Such experimental observations call for an in-depth theoretical analysis
of the effect of short-range correlations on Anderson localization.   

We describe briefly the theoretical investigations that have incorporated short-range as well as long-range spatial correlations in the diagonal as well as off-diagonal disorder. A series of one-dimensional versions of the Anderson model have been used to demonstrate a breakdown of Anderson’s localization driven by spatial correlations on the disorder distribution [11,12,13,14,15]. Also, effort has been made to demonstrate the strong effect of off-diagonal correlated disorder on Anderson localization. For example, a number of studies have employed correlated off-diagonal interactions and found delocalized states [16,17,18]. Besides short range correlations, several investigations have been performed considering long range correlations in the disorder distribution. Carpena et al~[19] find a long-range correlation-induced metal
-insulator transition using a one-dimensional tight-binding model. Francisco et al~[20] obtain an Anderson-like metal-insulator transition studying a one-dimensional tight-binding model with long-range correlated disorder. All these studies suggest that localization
properties are greatly renormalized when some kind of spatial correlation
is introduced in the disorder distribution. However, most of the studies are limited to electronic problems and Anderson localization of phonons in the presence of spatially correlated disorder has received scant attention, both theoretically and experimentally.

Being a general wave phenomenon, Anderson localization is ubiquitous. Sajeev John et al [21], using field theoretic techniques, investigated phonon localization in the presence of long range correlated random potential. However, methods like exact diagonalization (ED), transfer matrix method (TMM), multifractal analysis, diagrammatic techniques, itinerant coherent-potential approximation (ICPA) have not been employed for studying phonon localization in the presence of correlated disorder. Most of the mentioned methods have been confined to simple models of lattice vibrations, where the diagonal matrix elements $M(l)$ of the Hamiltonian are independent random variables.

In our previous study[22], we provided a detailed description of a typical medium dynamical cluster approximation (TMDCA), that yields a proper description of the Anderson localization transition in 3D. It adopts the typical density of states (TDOS) as a single particle order parameter for the Anderson localization transition (ALT) which makes it computationally less expensive compared to other numerical methods like ED and TMM. It satisfies all the essential requirements expected of a successful quantum cluster theory. We have also been able to extend the formalism for studying Anderson localization of phonons in the presence of both diagonal and off-diagonal disorder[22,23].

In this work, we investigate the nature of the Anderson transition for phonons in the presence of spatially correlated disorder in 3D. This paper is organized as follows. In section II, we give a brief description of the model and method that are used in this work. In section III, we present results and discussions. We conclude our work in section IV.

\section{Method}
As before [22], we consider the following Hamiltonian for the ionic degrees of freedom of a disordered lattice within the harmonic approximation in the momentum ($p$) and displacement($u$) basis, as
\begin{equation}
H=\sum_{\alpha i l}\frac{{p^2_{i\alpha}(l)}}{2M_{i}(l)} + \frac {1}{2} \sum_{\alpha\beta ll^\prime ij} \Phi^{\alpha\beta}_{ij}(l,l^\prime)u_{\alpha}^i(l)u_{\beta}^j(l^\prime)\,,
\label{mainham}
\end{equation}
where the symbols have their usual meaning as described in Ref [22]. In this work, we again restrict ourselves to a single branch ($\alpha$) and single basis atom ($i=1$) case, hence we drop the indices, $\alpha, \beta, i, j$. The unit cell index ($l$) is retained. The spatial dependence of the ionic masses $M(l)$ is incorporated through a local disorder potential $V$ as
\begin{equation}
{\hat{V}}_{ll^\prime} = \Big \lbrack 1 - M(l)/M
\Big\rbrack\delta_{l,l^\prime}\,.
\label{eq:dispot}
\end{equation}
In the previous work[22], we had considered a uniform box distribution, where the quantity $\Big(1-M(l)/M \Big)\in \lbrack -V, V \rbrack$ can take any value in that interval with equal probability and $0 \leq V \leq 1$ is the disorder strength. The random $V^{'}\text{s}$ from site to site were taken to be uncorrelated with each other. As mentioned in the introduction, the objective of this work is to investigate the effect of short-range correlations in the mass disorder.

We begin with nearest-neighbour correlations.
We first distribute masses randomly on the odd indexed sites and on the even indexed sites, exactly as was done previously, according to a uniform distribution with the same mean and variance. The disorder potential at the odd indexed sites is denoted as $V_1$ and that on the even indexed sites is denoted as $V_2$. Therefore, the following initial correlations hold:
\begin{equation}
\langle V_1 \rangle = 
\langle V_2 \rangle = 0\,.
\end{equation}
\begin{equation}
\langle V_1^2 \rangle = \sigma^2;\;
\langle V_2^2 \rangle = \sigma^2;\;
\langle V_1 V_2 \rangle = 0 \,.
\label{eq:variancedef}
\end{equation}
Now, since $V_1$ and $V_2$ are independent, $\rho_{V_1V_2}=0$. From these two uncorrelated random sequences, we want to generate correlations between consecutive sites of the odd and even sequences
with a specified correlation coefficient $\rho$.  The resulting new sequences for the odd and even indexed sites, denoted as $V_{\rm odd}$ and $V_{\rm even}$, should be correlated pairwise. So, the site $2n+1$ and $2n$ should be correlated. 
\begin{equation}
\rho_{V_{\text{odd}}V_{\text{even}}} = \frac{\big\langle \big(V_{\text{odd}} - \langle V_{\text{odd}}\rangle\big) \big( V_{\text{even}} - \langle V_{\text{even}}\rangle\big)\big\rangle}{\sigma^2} \ ,
\label{eq:corrdef}
\end{equation}
where $\sigma^2$ is the variance. Let us construct $V_{\text{odd}}$ and $V_{\text{even}}$ using linear combinations of $V_1$  and $V_2$ as
\begin{align}
& V_{\text{odd}} = a V_1 + b V_2 \nonumber\\
& V_{\text{even}} = c V_1 + d V_2 \,,
\label{eq:vrealtion}
\end{align}
where the unknown coefficients, $a,b,c$ and $d$ will be chosen so that
the odd and even sequences get correlated with each other.
So,
\begin{align}
\langle V_{\text{odd}} V_{\text{even}}\rangle & = \langle (aV_1 + b V _2) (c V _1 + d V _2)\rangle \nonumber \\
& = ac \langle V_1^2 \rangle + bd \langle V_2^2\rangle + (ad+bc) \langle V_1V_2 \rangle \ .
\label{eq:voddveven1}
\end{align}
Using Eq.\eqref{eq:variancedef} in Eq\eqref{eq:voddveven1}, we write
\begin{align}
\langle V_{\text{odd}} V_{\text{even}}\rangle = (ac + bd) \sigma^2 \ .
\label{eq:voddeven2}
\end{align}
Using Eq.\eqref{eq:variancedef} in Eq.\eqref{eq:corrdef}, we write
\begin{align}
\rho_{V_{\text{odd}}V_{\text{even}}} = \frac{\langle V_{\text{odd}} V_{\text{even}}\rangle}{\sigma^2} \ .
\label{eq:corrdef2}
\end{align}
Using Eq. \eqref{eq:voddeven2} in Eq. \eqref{eq:corrdef2}, we get
\begin{align}
\rho_{V_{\text{odd}}V_{\text{even}}} = ac + bd
\label{eq:corrdef3}
\end{align}
From Eq\eqref{eq:vrealtion}, we write
\begin{align}
\langle V_\text{odd} ^ 2\rangle & = \langle (a V_1 + b V_2) (a V_1 + b V_2)\rangle \nonumber \\
& = a^2 \langle V_1^2 \rangle + b^2 \langle V_2^2\rangle + (ac+bd) \langle V_1 V_2\rangle \ .
\end{align}
Using Eq.\eqref{eq:variancedef}, we get
\begin{align}
\langle V_\text{odd} ^ 2\rangle = (a^2+b^2)\sigma^2;
\langle V_{\text{even}}\rangle = (c^2 + d^2) \sigma^2 \,.
\label{eq:condvodd2}
\end{align}
We impose the condition
\begin{align}
\langle V_\text{odd}^2\rangle = \sigma^2; \langle V_\text{even}^2\rangle = \sigma^2 \,.
\label{eq:condvodd3}
\end{align}
From the above, it is easy to see that
\begin{align}
a^2+b^2 = c^2 + d^2 =1\,.
\label{eq:corrabcdnorm}
\end{align}
So, the transformation that yields the desired correlations can be chosen
as
\begin{align}
a = \cos{\phi}=d; b = \sin{\phi}=c\,.
\label{eq:corrabcdcal}
\end{align}
Hence, the expression
\begin{equation}
ac+bd = 2 \cos{\phi} \sin{\phi} = \sin{2\phi} \,.   
\end{equation}
Thus, random $V_{\text{odd}}$ and $V_{\text{even}}$ are correlated with 
$\rho_{V_{\text{odd}}V_{\text{even}}}$ which is equal to $\sin{2\phi}$, where 
\begin{equation}
\phi = \frac{1}{2}\sin^{-1}({\rho_{V_{\text{odd}}V_{\text{even}}}}) \, .
\label{eq:corrphical}
\end{equation}

\begin{figure}[!t]
\centering{
\includegraphics[width=1.0\columnwidth]{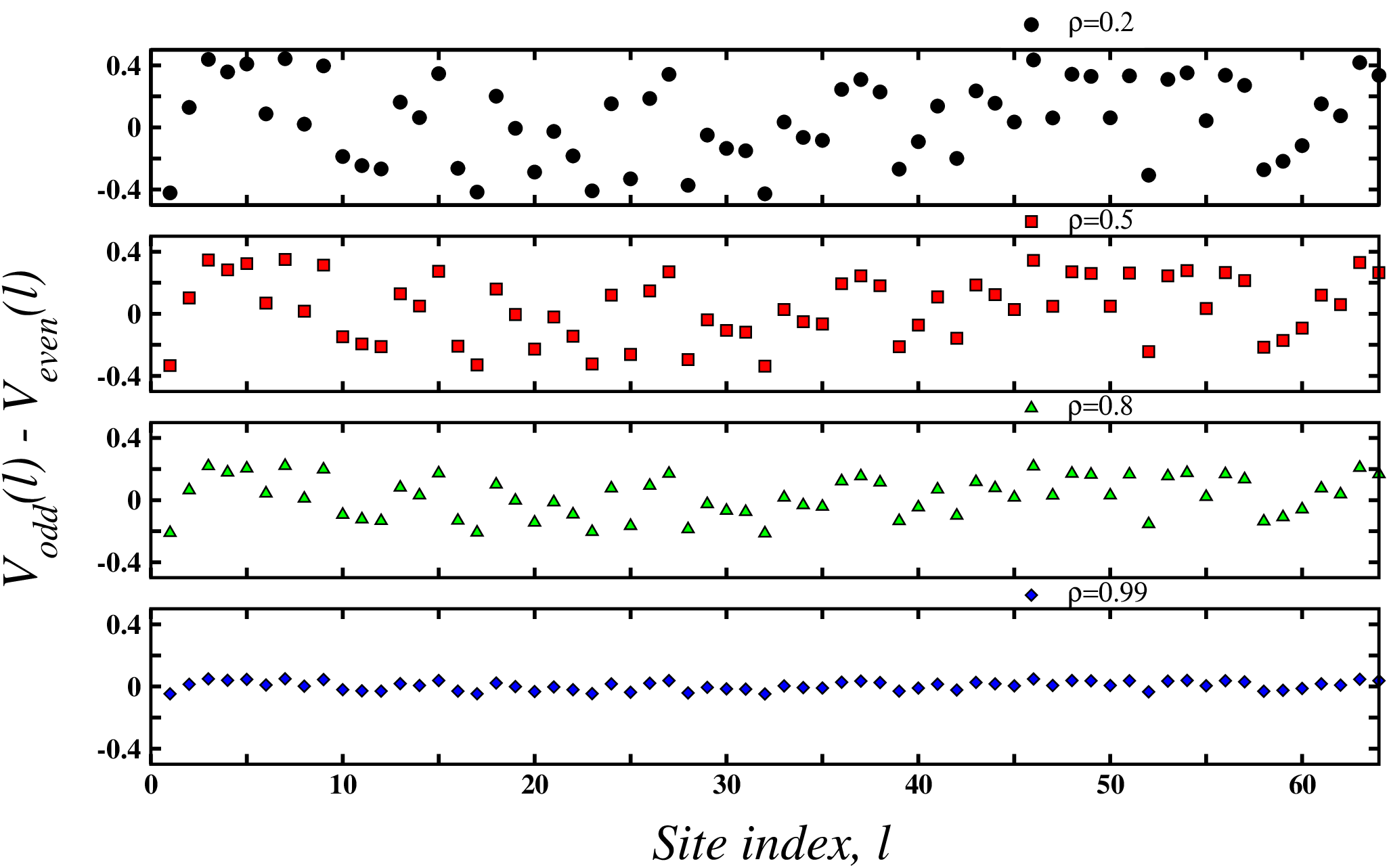}
\caption{A plot of the difference of  two correlated random sequences $V_{\text{odd}} - V_{\text{even}}$ for four different values of correlation coefficient ($\rho$) using cluster size $N_c=64$. Notice that the two random variables are strongly spatially correlated for $\rho=0.99$, whereas they are uncorrelated for $\rho=0.2$.}
\label{fig:scatteringplot}
}
\end{figure}

We can verify that this method does induce correlations between the even and the odd sequences. For vanishing correlation, i.e.\ for
${\rho_{V_{\text{odd}}V_{\text{even}}}}\rightarrow 0$, from
Eq.~\ref{eq:corrphical}, $\phi\rightarrow 0$ as well. This implies,
from Eqs.~\ref{eq:voddveven1} and ~\ref{eq:corrabcdcal}, that
$a,d\rightarrow 1$ and $b,c\rightarrow 0$, hence
\begin{align}
& V_{\text{odd}} \rightarrow V_1\nonumber\\
& V_{\text{even}} \rightarrow V_2 \ .
\label{eq:vrealtion_limit}
\end{align}
Since $V_1$ and $V_2$ are anyway uncorrelated, the new sequences,
$V_{\text{odd}}$ and $V_{\text{even}}$, in this limit are also uncorrelated. While in the other extreme, namely ${\rho_{V_{\text{odd}}V_{\text{even}}}}\rightarrow 1$, we get
$\phi\rightarrow \pi/4$, which implies $a,b,c,d\rightarrow 1/\sqrt{2}$, and hence $V_{\text{odd}}\simeq V_{\text{even}}\simeq(V_1+V_2)/\sqrt{2}$. Thus, in this limit,
$V_{\text{odd}}$ and $V_{\text{even}}$ become almost equal and are hence fully correlated. We illustrate this in Fig.~\ref{fig:scatteringplot}, where for four different correlation coefficients, $\rho_{V_{\text{odd}}V_{\text{even}}}=0.2, 0.5, 0.8$ and $0.99$, the difference of the two sequences, $V_{\text{odd}} - V_{\text{even}}$ is plotted as a function of the site-index. It is seen that for small correlation coefficients, the difference is large, and hence the odd and even sequences are uncorrelated. While for large correlation coefficient ($\gtrsim 0.9$), the difference is very small, and hence the two sequences are strongly correlated.

An  algorithm  that implements the described formalism for creating correlated disorder potential is stated below:

\noindent
1. The algorithm for generating correlated disorder potential starts with creating local disorder potential $V_l$, which we initially consider as spatially independent random variables distributed according to uniform (box) distribution as
\begin{equation}
P_{v}(V_l) = \Theta (V - |V_l|)/2V \ ,    
\end{equation}
where $V_l$ is the disorder potential defined in Eq.\eqref{eq:dispot} and $V$ is the width of the distribution that corresponds to the disorder strength. 

\noindent
2. Identify the $V_l$ at lattice sites $l$ that are labeled by the even number or odd number. We define $V_{1}(l)$ as the disorder potential at the odd indexed lattice sites and $V_{2}(l)$ as the disorder potential at the even indexed lattice sites.

\noindent
3. We set $\rho$ as correlation strength parameter which can be varied from 0 to 1. For a given value of $\rho$, we calculate $\phi$ using Eq.\eqref{eq:corrphical}. 

\noindent
4. The unknown coefficients $a, b, c, d$ are calculated using Eq.\ \ref{eq:corrabcdcal} and the normalization is maintained by imposing the condition given in Eq.\eqref{eq:corrabcdnorm}.  

\noindent
5. The spatial correlations among the $V_{\text{odd}}$ and $V_{\text{even}}$ are introduced depending on the strength $\rho$ according the relation given in Eq.\eqref{eq:vrealtion}.

The rest of the algorithm is the same as described in our previous publication[22].

\section{Results and discussion}
\begin{figure}[!t]
\centering{
\includegraphics[width=.6\columnwidth]{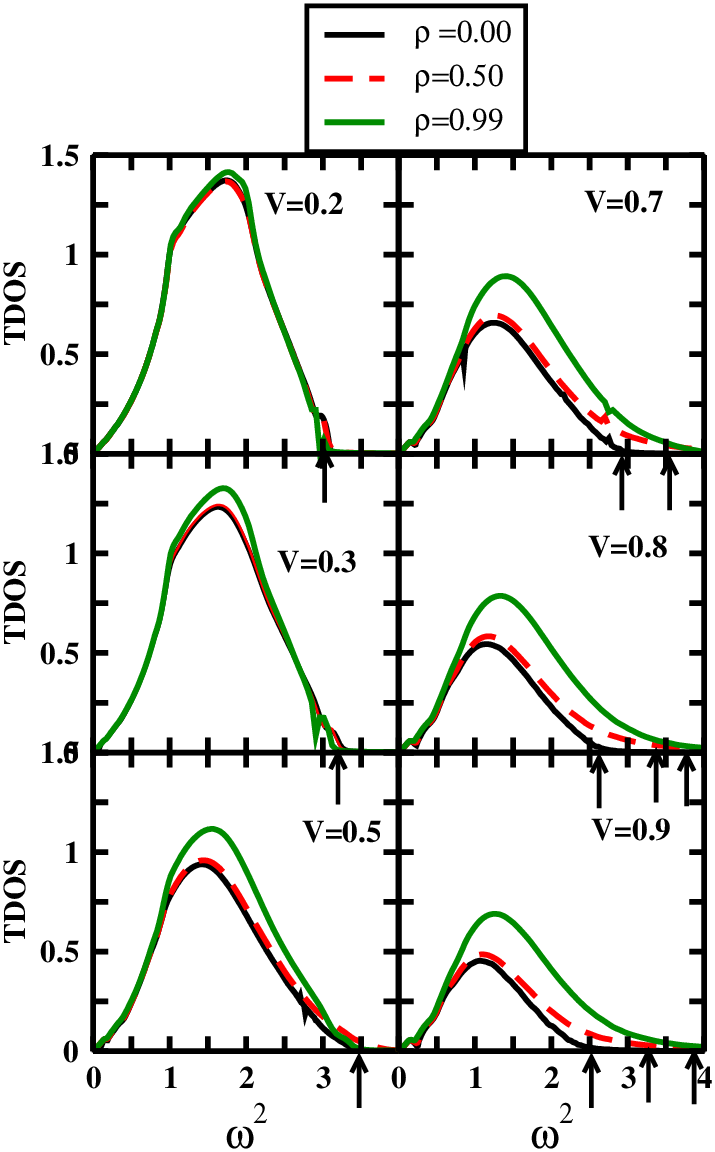}
\caption{The evolution of the TDOS, calculated from the TMDCA, as a function of the square of the frequency ($\omega^2$) with increasing disorder strength ($V$) considering a box distribution in three dimensions using cluster size $N_c=64$ for the uncorrelated ($\rho=0.00$) and correlated ($\rho=0.5, 0.99$) spatial disorder. The arrows shows the mobility edges.}
\label{fig:eval_TDOS_corr}
}
\end{figure}

\begin{figure}[!t]
\centering{
\includegraphics[width=.7\columnwidth]{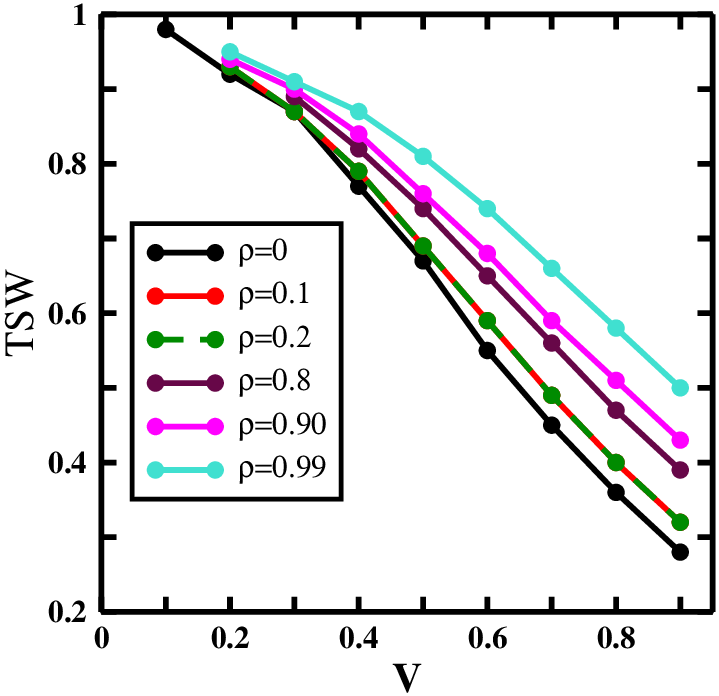}
\caption{Total spectral weight (TSW) of the TDOS as a function of increasing disorder strength ($V$) for correlated strength $\rho = 0.1-0.99$. We observe that the rate of decrease of the total spectral weight (TSW) of the TDOS decreases with increasing spatial correlation.}
\label{fig:spectral_TDOS_corr}
}
\end{figure}

\begin{figure}[!t]
\centering{
\includegraphics[width=.7\columnwidth]{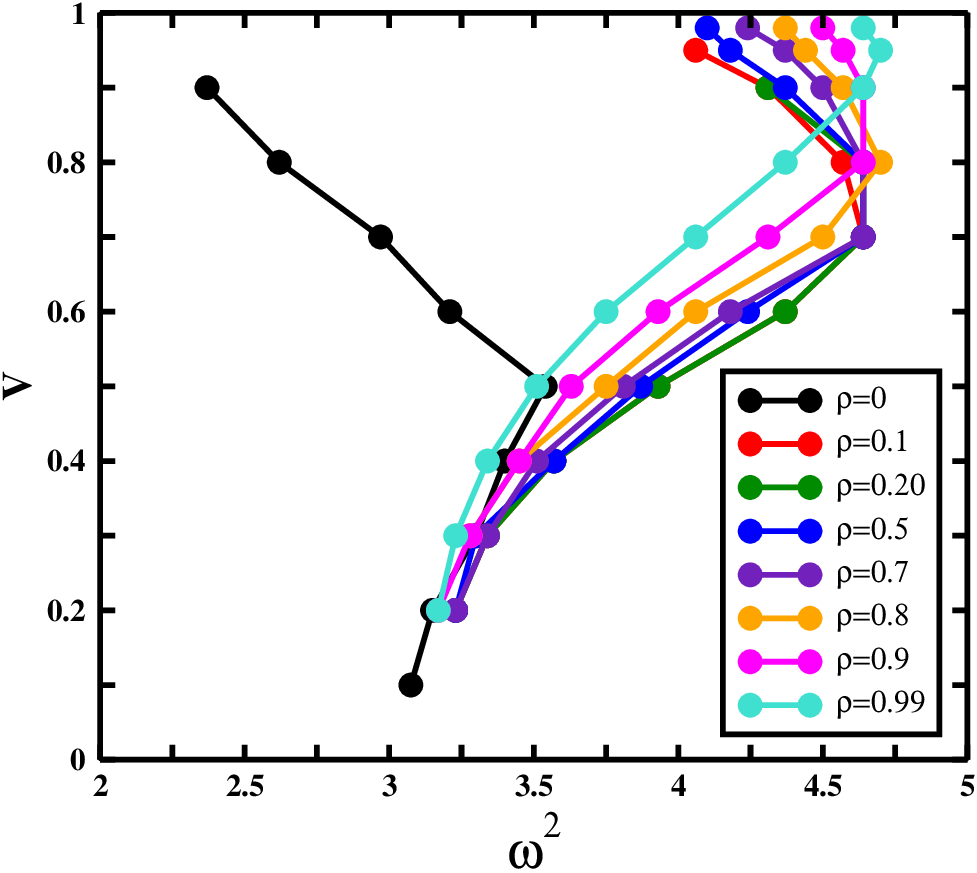}
}
\caption{Mobility edge trajectory for a box distribution of mass-disordered system in three dimensions. We find that the re-entrance behavior of the mobility edge shifts towards high-frequency region with intermediate correlation strength. Further increase of the correlation strength completely destroys the re-entrance behavior and mobility edges keeps on moving towards high-frequency region indicating a localization-delocalization transition driven by spatial correlations.}
\label{fig:mobility_edge_correlated}
\end{figure}

As we have already discussed, a true delocalization-localization transition occurs in 3D depending on the strength of disorder ($V$). We investigate this Anderson transition of phonons using the TMDCA in the presence of short range order. In our previous study [22], we have already established that the TDOS is a valid order parameter for studying phonon localization. So, we first observe the evolution of the TDOS with increasing disorder strength $V$ for correlated strength $\rho=0$ (uncorrelated) and $\rho=0.99$. It is displayed in Fig.\ref{fig:eval_TDOS_corr}. As may be expected, the TDOS for $\rho=0$ is almost the same as the TDOS for $\rho=0.99$ for low disorder ($V \leq 0.3$). But, for $V > 0.3$, the TDOS for $\rho=0.99$ starts to deviate strongly from the TDOS for the uncorrelated disorder. We note that the TDOS for $\rho=0.99$ differs significantly from the TDOS for the uncorrelated disorder at $V=0.9$. We have already understood that the vanishing of the TDOS implies the localization of vibrational modes[22]. Here we reproduce such behavior for $\rho=0$. The overall TDOS for $\rho=0$ 
decreases with increasing $V$ which indicates that the vibrational modes get localized as disorder increases. This kind of disorder-induced delocalization-localization transition is prevented by the introduction of spatial correlations in the system. Through a direct comparison of the TDOS for $\rho=0$ with the TDOS for $\rho=0.99$, such behavior can be easily observed. 
The mobility edges marked by the arrows represent the energy scale demarcating
the extended states from the localized states. Again, from Fig.\ref{fig:eval_TDOS_corr}, it is clear that the mobility edge shifts to higher energies with increasing correlation strength, thus implying that the latter induces delocalization of the hitherto localized states.

An alternative measure of the proximity to the Anderson localization transition is total spectral weight of the TDOS.  The variation of total spectral weight of the TDOS with increasing correlations is shown in Fig\ref{fig:spectral_TDOS_corr}. It clearly shows that the total spectral weight of the TDOS for $\rho=0.99$ decreases at a much slower rate compared to the uncorrelated disorder ($\rho=0$). Such behavior indicates that spatial correlations prevent the localization of vibrational modes. 

Another perspective of spatial correlations is obtained through an investigation of mobility edges which can be extracted from the TDOS presented in Fig\ref{fig:eval_TDOS_corr}. A mobility edge is defined as the energy which separates localized and extended  states[24]. The mobility edge has been measured for 3D Anderson localization[25]. The effects of spatial correlations on the mobility edges for Anderson localization of electrons have been studied extensively. However, to best of our knowledge, it has not been yet reported for the Anderson localization of phonons in the correlated disorder case. We define the mobility edge by the boundary of the TDOS and denote by arrows as indicated in Fig.\ref{fig:eval_TDOS_corr}.   

In Fig\ref{fig:mobility_edge_correlated}, we show calculated mobility edges using the TMDCA with $N_c=64$ for mass disorder. The phase diagram implicates that the spatially correlated diagonal disorder delocalizes the uncorrelated diagonal disorder induced localized vibrational modes. In the phase diagram, we first observe the usual behavior of the mobility edges with increasing $V$ for $\rho=0$. For small disorder $V < 0.5$, the trajectory of the mobility edges moves outward with increasing $V$. But, it starts moving inward for strong disorder $V \geq 0.5$. Thus, a re-entrance transition with increasing disorder occurs at $V=0.5$. We explored this behavior of the mobility edges in Ref[22]. The spatially correlated disorder destroys this re-entrant behavior of the mobility edges. As seen in Fig\ref{fig:mobility_edge_correlated}, the trajectory of the mobility edges for $\rho=0.99$ is almost the same as that for $\rho=0$ in the presence of small disorder $V \leq 0.5$. However, in contrast to the uncorrelated case, the trajectory of the mobility edges keeps on moving outward with increasing disorder strength $V > 0.5$. It suggests that the spatial correlations drive the system towards delocalization.  


\section{Conclusions}\balance
We have applied the TMDCA formalism for investigating the effects of short-range spatial correlations on phonon localization in 3D. 
We have only considered pairwise correlations between the adjacent odd-indexed and even-indexed sites. The correlation strength is varied from 0 to 1. In the weak correlation limit, all the sites have completely random masses, while in the strong correlation limit, the masses of the $(2l-1)^{\rm th}$ site and the $(2l)^{\rm th}$ site are the same, but as a function of $l$, the odd/even sequence of masses is still random. Our main conclusion is that correlated disorder with just pairwise correlations can markedly change the localization transition of phonons.
Such a conclusion is validated by observing the variation of the TDOS and mobility edges with increasing correlation strength. We show that short-range correlated disorder impedes the localization of the vibrational modes, and eventually, a correlation induced localization-delocalization transition of phonons sets in a 3D disordered sample. It would certainly be valuable to understand the observed delocalization transition in the presence of long-range correlated disorder. For doing so, an extension of the current framework incorporating long-range correlations is in progress. 


\section*{Acknowledgement}


\end{document}